# Probabilistic computation of wind farm power generation based on wind turbine dynamic modeling

Herman Bayem, Yannick Phulpin, Philippe Dessante and Julien Bect

*Abstract*—This paper addresses the problem of predicting a wind farm's power generation when no or few statistical data is available. The study is based on a time-series wind speed model and on a simple dynamic model of a doubly-fed induction generator wind turbine including cut-off and cut-in behaviours. The wind turbine is modeled as a stochastic hybrid system with three operation modes. Numerical results, obtained using Monte-Carlo simulations, provide the annual distribution of a wind farm's active power generation. For different numbers of wind turbines, we compare the numerical results obtained using the dynamic model with those obtained considering the wind turbine's steady-state power curve. Simulations show that considering the wind turbine's dynamics does not significantly enhance the accuracy of the annual distribution of a wind farm generation.

*Index Terms*-- wind farm, stochastic hybrid system, cut-off behaviour

## I. INTRODUCTION

NUMEROUS reports emphasize that operation and planning in modern power systems are highly impacted by wind power generation [1]-[3]. The significant integration of wind turbines requires thus utilities to develop new tools for power system management. Actually, though statistical data can be collected by system operators, probabilistic tools are needed in order to achieve a more efficient use of the system capabilities and better planning for new investments. For example, several recent works have focused on probabilistic load-flow computation [4]-[5] in order to ease operational tasks for transmission system operators. As a load-flow computation requires the knowledge of every injection and demand, each wind turbine output power, which is naturally volatile, should be accounted for through its probability density function (pdf) in those applications [6]-[7].

There exist several approaches for assessing the probability density function of wind turbine power generation. One could take advantage of the progress in wind forecasting [8] and eventually assess the scheduled pdf of a wind turbine's power injection. System operators could then schedule power system operation depending on the forecasted hourly average wind speed. However, this approach seems difficult to implement since wind speed forecasts are highly uncertain. On the other hand, one could consider, as in [6] or [9], the annual distribution of a wind turbine's power generation based on its steady-state power curve and using annual wind speed distributions. However, this approach does not consider the wind turbine's dynamics, which may impact its active power injection [10]-[11].

Our approach aims to analyze whether the consideration of a wind turbine's dynamics impacts the annual distribution of its power generation. The study is based on a time-series model for the wind speed. While an appropriate model should be designed for every type of generator, we will mainly focus in this paper on a Doubly-Fed Induction Generator (DFIG) wind turbine which will be modelled as a stochastic hybrid system with three operation modes. This dynamic modeling includes cut-off and cut-in behaviours, which can have a great influence on the wind turbine's generation availability in certain wind speed ranges [12]. Finally, numerical results, obtained using Monte-Carlo simulations, provide the annual distribution of the active power generation for a single wind turbine and for a wind farm with several wind turbines. Those numerical results are compared with the annual distribution obtained using the steady-state power curve of the wind turbine.

This paper is organised as follows: in Section II, the wind model is detailed. Section III describes the wind turbine while Section IV defines the three mode control scheme used in our simulations. Finally, the simulation process, the numerical data and the simulation results are presented in Section V.

## II. WIND MODEL

As introduced in [13], we consider a "wide band" model for the horizontal wind speed $v(t)$ of the form:

$$v(t) = \bar{v}(t) + w(t) \qquad (1)$$

where $\bar{v}(t)$ is a slowly varying signal, modeling the hourly wind speed variations, and $w(t)$ a highly fluctuating signal, modeling the high frequency turbulent phenomena. Moreover, conditionally to $\bar{v}(t) = \bar{v}$, the fast component $w(t)$ is modeled as a stationary Gaussian process [13]-[14] with mean $\bar{v}$ and standard deviation $\sigma = \kappa \cdot \bar{v}$ (with $\kappa$ a positive constant).

We use the same model as in [11] and [14] for the dynamics of the fast component:

H. Bayem, Y. Phulpin and P. Dessante are with the Department of Power and Energy Systems in SUPELEC, Paris, France (e-mail: {herman.bayem, yannick.phulpin, philippe.dessante}@supelec.fr).

J. Bect is with the Department of Signal Processing and Electronic Systems in SUPELEC, Paris, France (e-mail: julien.bect@supelec.fr).

$$\frac{dw}{dt}(t) = -\frac{w(t)}{T} + \kappa \overline{v}(t)\sqrt{2/T}\xi(t) \qquad (2)$$

where $\xi(t)$ is a Gaussian white noise (formal derivative of standard Brownian motion) and $T = L/\overline{v}$, with $L$ the turbulence length scale. Experimental values of both $\kappa$ and $L$ can be found in [14], for example.

The slow component $\overline{v}(t)$ is inspired by the ARMA(3,2) process used in [15]. The modifications are a "reflection" at $\overline{v} = 0$, introduced to keep the time series positive at all times, and a linear interpolation to convert the time series into a continuous-time process. For numerical purposes, we will use the ARMA parameters given in [15] for the "Swift Current" location.

The time-series model of the horizontal wind speed $v(t)$ is thus completely specified. Although extremely simplified in several respects (long-time correlations are obviously underestimated by the ARMA model, for instance), it should be sufficient for the purpose of this paper. We refer to [11] for a more detailed discussion concerning the model for the fast component.

III. WIND TURBINE MODEL

A wind turbine model is generally composed of three subsystems representing its main components, namely the rotor, the generator, and the gearbox, which connects the rotor and generator shafts. In this paper, we use a variable speed wind turbine, whose generator is connected to the grid through a power electronic converter. This type of wind turbine is generally equipped with controllers which allow the control of the wind turbine's pitch angle and active/reactive power output. While an exhaustive description of such a wind turbine can be found in [13], we present hereafter a simple dynamic model.

*A. General representation of a variable speed DFIG wind turbine*

We consider a variable speed pitch controlled wind turbine with a doubly fed induction generator. As described in [13], the stator of the DFIG is directly connected to the grid while its rotor is coupled to the grid through a power converter, which allows controlling the active and reactive power output. This scheme is presented in Figure 1.

The grid is represented by a slack bus, into which the wind turbine's active power generation is injected at any time. In this context, the dynamics of the wind turbine related to the variations of the wind speed are controlled trough three controllers, namely the rotor speed controller, the pitch angle controller and the state controller. Figure 2 shows how those elements (rotor, generator and controllers) interact. Section III-B details the different elements of the wind turbine while the control scheme is presented in section IV.

*B. Wind turbine model*

*1) Turbine model*

The mechanical power captured by the turbine is given by the following equation:

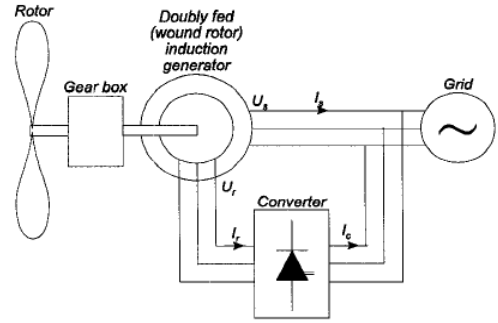

Figure 1: Connection of a DFIG wind turbine to the electrical network [16].

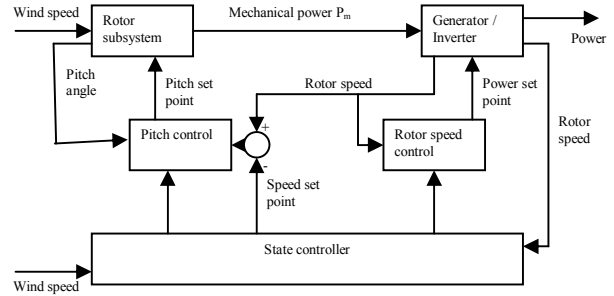

Figure 2: interactions between wind turbine's elements.

$$P_m(v,\theta,\omega) = \frac{\pi}{2} \cdot \rho \cdot R^2 \cdot C_P(\lambda,\theta) \cdot v^3 \qquad (3)$$

where $v$ is the wind speed, $\omega$ the wind turbine rotational speed, $\rho$ is the air density, $R$ the turbine radius, $\theta$ the pitch angle, $\lambda$ the tip-speed ratio ($\lambda = R.\omega/v$) and $C_p$ is the power coefficient, which describes the turbine's ability to convert wind kinetic energy into mechanical energy.

*2) Generator drive train model*

The relation between driving and braking torques and the acceleration of the turbine can be expressed as follows:

$$J \cdot \dot{\omega} = \frac{P_m - P_g}{\omega} \qquad (4)$$

where $J$ is the total inertia of the system, $P_m$ is the mechanical power that resulting from the force received by the rotor, and $P_g$ is the electro-mechanical power exercised by the DFIG. For the sake of simplicity, we do not consider friction torque in this study.

*3) DFIG model*

We consider a generator efficiency $\eta$ such that the active power injected in the grid $P_{g,out}$ is equal to 90% of the electro-mechanical power $P_g$.

*4) Rotor Speed controller*

The rotor speed controller sets the active power set point $P_g$ according to the rotor speed. $P_g$ also depends on the functioning modes of the wind turbine, which are detailed in Section IV.

*5) Pitch controller*

As represented in Figure 3, the power coefficient $C_p$ is dependent on the tip-speed ratio $\lambda$ and on the pitch angle $\theta$.

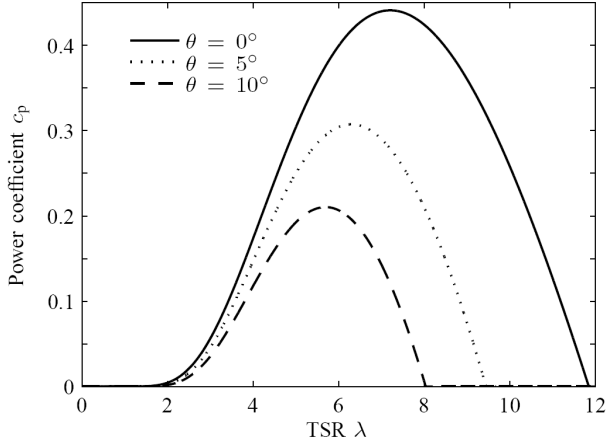

Figure 3: Power coefficient as a function of the tip speed ratio for different pitch angle.

When the turbine is generating, the pitch angle controller limits the mechanical power captured by the turbine in order to maintain the rotor speed within acceptable limits. As in [16], it is modeled as a proportional controller with saturation limits on the pitch rate:

$$\dot{\theta} = \begin{cases} 0 & \text{if } \theta = 0 \text{ and } \omega \leq \omega_{nom} \\ 0 & \text{if } \theta = \theta_{max} \text{ and } \omega \geq \omega_{nom} \\ h(K_2 \cdot (\omega - \omega_{nom})) & \text{otherwise} \end{cases} \quad (5)$$

with:

$$h(\dot{\theta}) = \min(\dot{\theta}_{max}, \max(\dot{\theta}_{min}, \dot{\theta})). \quad (6)$$

*6) State controller*

In order to operate the DFIG wind turbine in an acceptable speed range, the state controller enables the pitch control and the rotor speed control to operate depending on the wind speed and the rotor speed as described in section IV.

## IV. CONTROL SCHEME FOR THE WIND TURBINE

In order to ease the understanding of the wind turbine's dynamics with a time-varying wind speed, we use a simple control scheme with three modes. All types of operation are covered, including cut-off and cut-in behaviours. Figure 4 summarizes the control scheme of this hybrid system, which is detailed in this section.

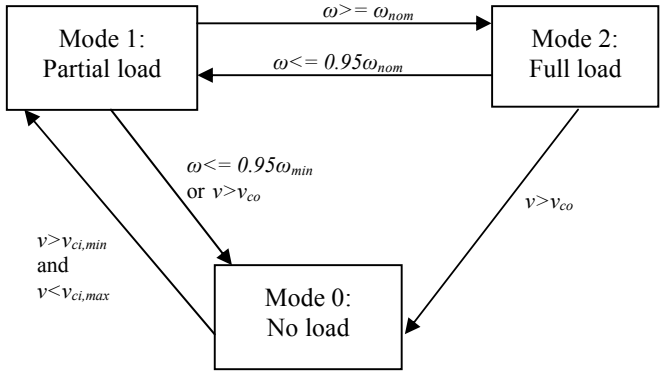

Figure 4: Control scheme of the variable speed DFIG wind turbine.

In the above figure, the condition $v > v_{co}$ means:

$$\bar{v}_5 > v_5^{cut-off} \quad or \quad \bar{v}_{60} > v_{60}^{cut-off} \quad (7)$$

where $\bar{v}_5$ and $\bar{v}_{60}$ are the average wind speed over the last 5 and 60 seconds, respectively, $v_5^{cut-in}$ is the cut-in wind speed and $v_5^{cut-off}$ and $v_{60}^{cut-off}$ the short term and long term cut-off wind speeds, respectively. The conditions given in equation (7) were inspired by [17].

*A. Mode 0: no load*

"Mode 0" corresponds to situations where the turbine does not generate any electrical power. This situation occurs when the turbine is completely stopped due to low or high wind speeds or due to a manual stop. By extension, we also consider that the wind turbine is in this mode during starting phases, when the generator is not yet coupled to the grid. Since we are concerned about the annual distribution of the power generation, this extension should have a low impact on the simulation results.

Transition from "mode 0" is activated when the average wind speed meets the "normal functioning" requirements, which are deduced from [17] and defined as follows:

$$\begin{cases} \bar{v}_{60} \geq v_{60}^{cut-in} \\ \bar{v}_5 \leq v_5^{cut-off} \\ \bar{v}_{60} \leq v_{60}^{cut-off} \end{cases} \quad (8)$$

In this mode, the power demand is set to zero ($P_g = 0$), and neither $\omega$ nor $\theta$ are explicitly modeled (since they are not needed). When the transition is activated, the wind turbine is switched to mode 1 with $\omega = \omega_{min}$ and $\theta = 0$.

*B. Mode 1: partial load*

For medium wind speeds, the rotor speed is controlled by setting the generator output power according to the following equation inspired by [18]:

$$P_g(\omega) = \frac{\pi}{2} \rho \cdot R^2 \cdot C_p \left( \frac{R\omega}{v_1(\omega)}, 0 \right) \cdot v_1(\omega)^3, \quad (9)$$

with:

$$v_1(\omega) = \frac{\omega - \omega_{min}}{\omega_{nom} - \omega_{min}} (v_{nom} - v_{ci}) + v_{ci} \quad (10)$$

where $\omega_{nom}$ is the nominal value for the rotor speed and $v_{nom}$ the wind speed for which this rotor speed is achieved under steady-state conditions.

Two transitions from "mode 1" can be activated:
- when the rotor speed $\omega$ becomes lower than $0.95\omega_{min}$ or the average wind speed becomes higher than the cut-off wind speed ($v > v_{c-o}$), then the wind turbine is switched to "mode 0".
- when the rotational speed $\omega$ becomes higher than $\omega_{nom}$, defined as the rotation speed for which $P_g$ reaches its nominal value $P_{nom}$, then the wind turbine is switched to "mode 2".

*C. Mode 2: full load*

"Mode 2" corresponds to higher wind speeds for which the rotor speed is close to $\omega_{nom}$ and the pitch controller is activated in order to limit the mechanical power and to maintain the

output power around its nominal value. The generator power reference is:

$$P_g(\omega) = \frac{\omega}{\omega_{nom}} P_{g,nom} \quad (11)$$

where $P_{g,nom}$ is the rated power of the wind turbine.

Two transitions can be activated from mode 2:
- when the average wind speed becomes higher than the cut-out wind speed ($v > v_{co}$), then the wind turbine is switched to "mode 0".
- when the rotational speed $\omega$ becomes lower than $0,95 \cdot \omega_{nom}$, the wind turbine is switched to "mode 1".

## V. SIMULATION RESULTS

### A. Simulation process

The wind and wind turbine models under consideration are relevant for simulating the power production of a wind farm in a large time-scale (months or years). For this study, the time-scale is one year.

The first part of the simulation process is the Monte-Carlo sampling of hourly mean wind speed values according to the probabilistic wind model described in Section II. Then, the power production of the wind turbine is computed according to its dynamic model and to its steady-state power curve, respectively.

When the wind farm is composed of several wind turbines, each turbine $i$ is supposed to receive a wind with the same slowly varying component $\bar{v}(t)$ and an independent fluctuating signal $w_i(t)$.

### B. Numerical data

Numerical data for the wind turbine and the wind speed model are presented in Table I and Table II, respectively. Two values (*5,46* m/s and *10,00* m/s) have been chosen for the annual mean wind speed value.

TABLE I
WIND TURBINE FEATURES.

|  | Name | Value |
| --- | --- | --- |
| $R$ | Rotor radius | 37.5m |
| $\omega_{min}$ | Minimal rotor speed | 9 RPM |
| $\omega_{nom}$ | Nominal rotor speed | 18 RPM |
| $P_{g,nom}$ | Nominal power | 2,03 MW |
| $v_{nom}$ | Nominal wind speed | 14 m/s |
| $v_{ci,min}$ | Cut-in wind speed | 3.5 m/s |
| $v_{ci,max}$ | Restart wind speed (after cut-off) | 19 m/s |
| $v_5^{cut-off}$ | "Fast" cut-out speed | 25 m/s |
| $v_{60}^{cut-off}$ | "Slow" cut-out speed | 20 m/s |
| $J$ | Turbine inertia | $1,4 \cdot 10^6$ kg.m$^2$ |

TABLE II
WIND MODEL FEATURES.

|  | Name | Value |
| --- | --- | --- |
| $\rho$ | Air density | 1.134 kg/m$^3$ |
| $L$ | Turbulence length scale | 300 m |
| $\kappa$ | Std / mean ratio for $\bar{v}$ | 0.15 |

### C. Single wind turbine generation distribution

The simulation process was first applied to a single wind turbine with two different wind speed time series (with an annual mean wind speed of *5.46* m/s and *10* m/s respectively). The joint probability function of the wind turbine power production and obtained wind speed is shown in Figure 5 and Figure 6.

One can notice that the distributions of the wind turbine power generation as a function of the instantaneous wind speed that is represented in Figure 5 is similar to the experimental distribution reported in [19]. It can also be observed on Figures 5 and 6 that, for certain instantaneous values of the wind speed, there may be a considerable uncertainty in the power production in comparison to the steady-state power. This uncertainty is even more significant when considering sites with higher annual mean wind speed since the cut-off behaviour of the wind turbine is poorly modeled by the steady-state power curve.

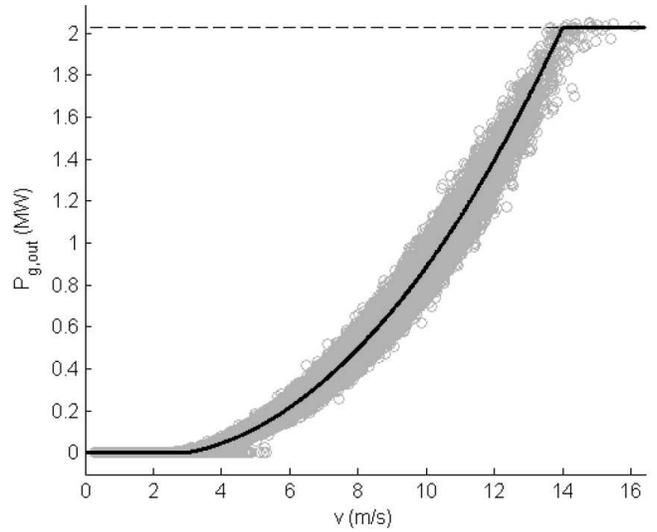

Figure 5: Power output obtained from the dynamic model (grey) for a wind time-series of *5.46* m/s annual mean. The steady-state power curve is represented in black.

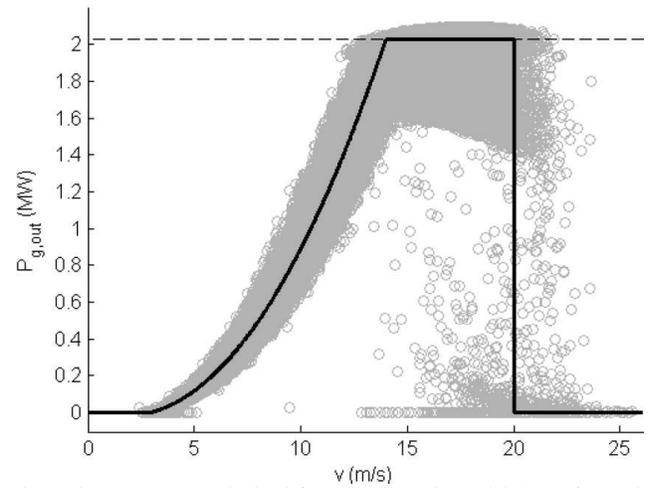

Figure 6: Power output obtained from the dynamic model (grey) for a wind time series of *10.00* m/s annual mean. The steady-state power curve is represented in black.

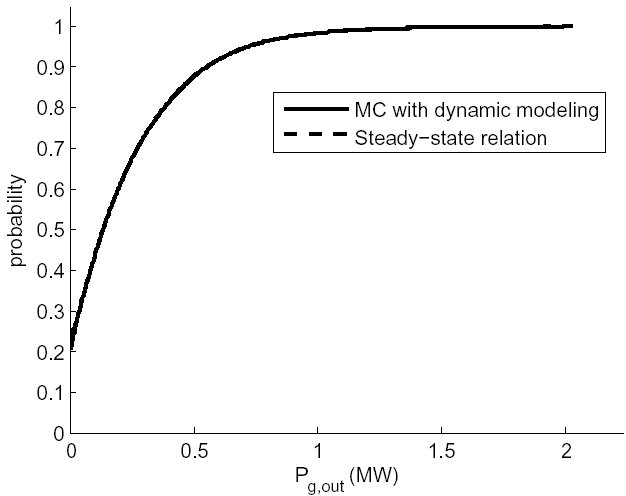

Figure 7: Cumulative distribution function of the wind turbine generation obtained with dynamic modeling and steady-state power curve with an annual mean wind speed of *5.46* m/s.

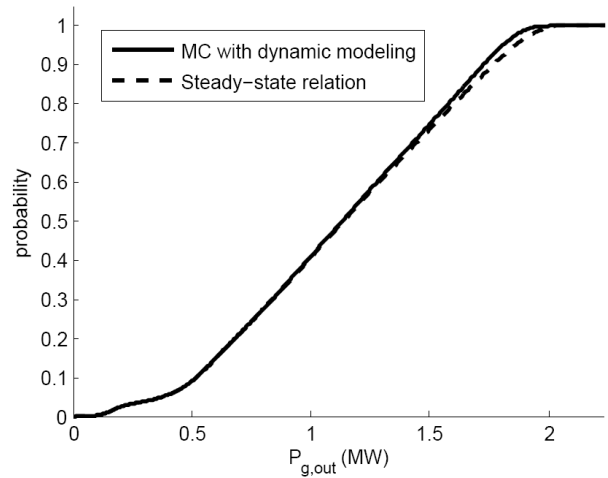

Figure 9: Cumulative distribution function of a 5 turbine wind farm generation obtained with dynamic modeling and steady-state power curve with an annual mean wind speed of *10* m/s.

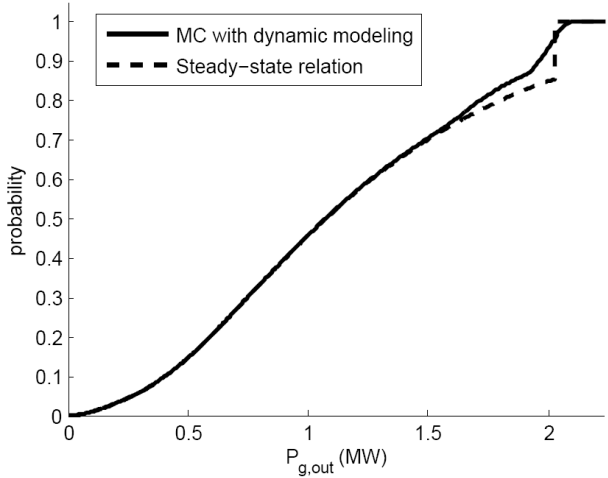

Figure 8: Cumulative distribution function of the wind turbine generation obtained with dynamic modeling and steady-state power curve with an annual mean wind speed of *10* m/s.

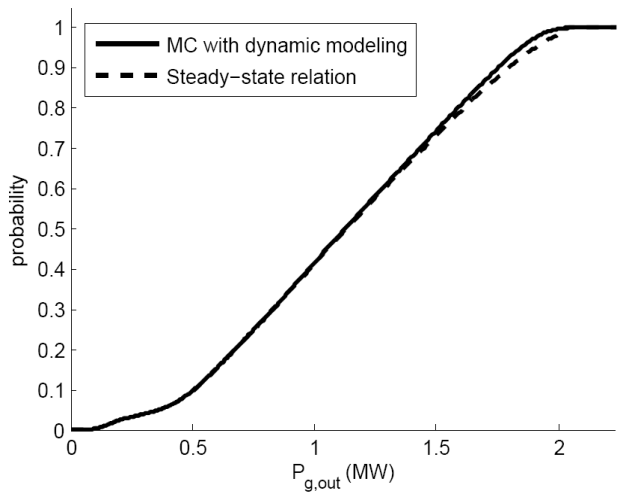

Figure 10: Cumulative distribution function of a 10 turbine wind farm generation obtained with dynamic modeling and steady-state power curve with an annual mean wind speed of *10* m/s.

For a low annual mean wind speed, similar results are observed when computing the cumulative distribution function of the wind turbine power production with dynamic modeling and with the steady-state power curve of the wind turbine (Figure 7). This observation may be explained by a lower uncertainty when it is computed over a long period of time (i.e. one year instead of one second). A slight difference is remarkable when considering a site with higher annual mean wind speed (Figure 8). In this case, the dynamic modeling of the wind turbine with high winds and the cut-off consideration may indeed be significant.

On Figure 8, it can be observed that the probability to have no power output is close to zero. Although this may be surprising with regard to the three mode scheme previously designed, this result is mainly caused by the wind model (mainly based on the average wind speed), which rarely generates wind speed values lower than 2 or 3 m/s for an average speed of 10 m/s. Nevertheless, this observation emphasizes the necessity of developing a model more adequate to the simulation of long-term wind variations.

### D. Wind farm generation distribution

Simulations were also performed for wind farms with 5 and 10 turbines, respectively. As emphasized for Figure 7, the annual distribution of a wind farm on a site with *5.46* m/s annual mean wind speed are similar when considering the steady-state power curve or the dynamic model for the wind turbines.

Figures 9 and 10 represent the power generation of the wind farm ($P_{g,out}$) normalized by the farm size (5 and 10 turbines, respectively) for a site with *10* m/s annual mean wind speed. Differences between the two wind turbine models are particularly small. As introduced for Figure 8, small wind values are particularly unexpected with such an annual mean wind speed. Consequently, the probability of no production is close to zero.

## VI. Conclusion

In this paper, Monte-Carlo sampling is performed on a probabilistic wind model to obtain a wind speed time series for one year. The resulting distribution is combined with a dynamic model of a wind turbine in order to compute its generation distribution. For comparison, this distribution is also applied to the steady-state power curve of the wind turbine.

Simulations lead to small differences between those two models, which are mainly observed for high wind speeds, close to the cut-off values. Moreover, the probabilistic computation of a multi-turbine wind farm using a correlation between wind speeds for each turbine shows that this difference may decrease when considering sites with several wind turbines.

This paper thus emphasizes that the dynamics of the wind turbines are not significant with respect to the annual distribution of wind farm power generation. This finding justifies the consideration of the steady-state power curve for some specific probabilistic Load-Flow computation, for example.

However, the obtained wind speed distribution is certainly dependent on the wind model. Further investigations in wind speed time-series modeling are thus required in order to achieve a better probabilistic modeling of a wind farm's power generation.